\documentclass[journal]{IEEEtran}

\ifCLASSINFOpdf
\else
\fi

\usepackage{tabularx}
\usepackage{array,multirow,graphicx}
\usepackage[table]{xcolor}
\usepackage{makecell}
\usepackage{placeins}
\usepackage{amsmath, amssymb}
\usepackage{verbatim}
\usepackage{enumitem}
\usepackage{float}
\usepackage[T1]{fontenc}
\usepackage[utf8]{inputenc}
\usepackage{frcursive}
\usepackage{fancyvrb}
\usepackage{fvextra}
\usepackage{graphicx}
\usepackage{hyperref}
\usepackage{url}
\usepackage[normalem]{ulem}
\graphicspath{{Figures/}}
\begin{document}
\title{STRisk: A Socio-Technical Approach to Assess Hacking Breaches Risk}

\author{Hicham Hammouchi, \IEEEmembership{IEEE Member,}
        Narjisse Nejjari, 
        Ghita Mezzour, 
        Mounir Ghogho, \IEEEmembership{IEEE Fellow,}
        Houda Benbrahim

\thanks{This work is partly funded by the NATO Science for Peace and Security (SPS) programme under research contract SPS G5319 "Threat Predict (\href{https://threatpredict.inria.fr/}{https://threatpredict.inria.fr/}): From Global Social and Technical Big Data to Cyber Threat Forecast"}

\thanks{H. Hammouchi, N. Nejjari, G. Mezzour, and M. Ghogho are with the College of Engineering \& Architecture (TICLab), Université Internationale de Rabat, Rabat 11100 Morocco (e-mail: hicham.hammouchi@uir.ac.ma; ghita.mezzour@uir.ac.ma; narjisse.nejjari@uir.ac.ma; mounir.ghogho@uir.ac.ma)}
\thanks{H. Hammouchi, N. Nejjari and H. Benbrahim are with ENSIAS, Université Mohammed V, Rabat 10110 Morocco (e-mail: hicham.hammouchi@uir.ac.ma; narjisse.nejjari@uir.ac.ma; benbrahimh@hotmail.com)}
}


\maketitle
\begin{abstract}


Data breaches have begun to take on new dimensions and their prediction is becoming of great importance to organizations. Prior work has addressed this issue mainly from a technical perspective and neglected other interfering aspects such as the social media dimension. To fill this gap, we propose STRisk which is a predictive system where we expand the scope of the prediction task by bringing into play the social media dimension. We study over 3800 US organizations including both victim and non-victim organizations. For each organization, we design a profile composed of a variety of externally measured technical indicators and social factors. In addition, to account for unreported incidents, we consider the non-victim sample to be noisy and propose a noise correction approach to correct mislabeled organizations. We then build several machine learning models to predict whether an organization is exposed to experience a hacking breach. By exploiting both technical and social features, we achieve a Area Under Curve (AUC) score exceeding 98\%, which is 12\% higher than the AUC achieved using only technical features. Furthermore, our feature importance analysis reveals that open ports and expired certificates are the best technical predictors, while spreadability and agreeability are the best social predictors.


\end{abstract}

\begin{IEEEkeywords}
Data breach, hacking breach, Twitter, cyber risk assessment, cyber risk prediction
\end{IEEEkeywords}

\IEEEpeerreviewmaketitle

\section{Introduction}

\IEEEPARstart{D}{ata breaches} have plagued both large and small organizations for years. In 2019 alone, 7,098 data breaches were reported, allegedly disclosing more than 15 billion sensitive records, an increase of 284\% over the previous year. Of these data breaches, 5,184 (72.5\%) were caused by hacking alone, resulting in the breach of more than 1.5 billion records \cite{QuickViewreport2019}. These breaches result in exfiltrating sensitive information including identity theft, social security numbers, credit card credentials, and medical records. Once this information is stolen, it becomes accessible in darkweb market places, where hackers sell the stolen data, share and sell tools and techniques to conduct attacks. Despite the high impact and business damage of these breaches, organizations still suffer from a long detection delay which can take up to 206 days on average. Besides, the containment of the breach may take up to 73 more days on average, totaling 279 days from the launch of the attack to its containment \cite{IBMBreachCost2019}. This gives attackers ample time to sell the breached information on darkweb forums and social media platforms, while victims are unaware of the breach. It is therefore important for an organization to assess its risk of exposure from various perspectives and to predict its exposure to experience a data breach.

Recently, proactive approaches to  cybersecurity have attracted much attention from the research community. A key element of these approaches is the the prediction of cyber-attacks. In the context of data breaches, the existing work on  prediction analysis ranges from external assessment of organizations' security posture based on network mismanagements and IP intelligence \cite{liu2015cloudy, sarabi2016risky}, to estimating the probability of a breach to occur based on the study of past incidents and their intrinsic statistical properties \cite{xu2018modeling}. 
The existing work, however, overlooked the interaction between breach incidents and other contextual aspects such as the social media dimension. In fact, taking these aspects into account can lead to better understanding and prediction of data breach incidents. It is worth pointing out that social media were used in previous work to detect and understand certain types of cyber threats but not data breaches. For instance, Twitter was considered to extract security events such as DDoS and data breaches \cite{ritter2015weakly}; and darkweb discussions combined with Twitter were used to predict some types of attacks \cite{goyal2018discovering}.

In this paper, we present STRisk, a predictive system in which we analyze the technical security posture and social reputation on Twitter of more than 3800 US organizations in order to predict their exposure to experience a data breach caused by hacking. We focus our predicting hacking breaches caused by hacking activities and malwares since these two types of breaches are the most prevalent and the most damaging ones \cite{dbirverizon19, hammouchi2019digging}. To explore the potential of STRisk, we first collect ground truth data consisting of victim organizations from two projects for reporting publicly disclosed data breaches, namely Privacy Rights Clearinghouse (PRC) and Veris Community Database (VCDB). For the non-victim (negative) sample, we select randomly from the American Registry for Internet Numbers (ARIN) a set of organizations that did not report a hacking breach. We then build a socio-technical profile for each organization by leveraging a variety of indicators capturing both technical anomalies and social signals. Technical anomalies include blacklisted hosts, open ports, expired web certificates, and spam activities, while social signals obtained from Twitter consist of tweets' sentiments, popularity, spreadabiltiy and other indicators. We build these profiles to bring insights into the digital reputation and technical posture of an organization as seen from outside. 

Before predicting organizations' exposure to data breaches, we handle the noise in the dataset by detecting and correcting mislabeled organizations, as many organizations in the negative sample may have experienced a breach and did not report it. This problem is known as noisy labels in the machine learning literature; the challenge is learn predictive models in the presence of noisy labels. We propose a method to identify and correct mislabeled non-victim organizations, and then  train STRisk by building various supervised machine learning models including tree-based models such as boosting and bagging models, and linear-based models such as logistic regression and Support Vector Machines (SVM). We train these models to distinguish between victim and non-victim organizations, and to assess the extent to which the designed indicators can inform which organizations are at risk of falling victim to a data breach.

In summary, in this paper we make the following contributions:
\begin{enumerate}[label=(\alph*)]

    \item We collect a variety of datasets and we design a set of features capturing both technical indicators, covering network anomalies and malicious activities, and social signals collected from Twitter. 
    
    \item We account for noisy labels in the negative sample and we propose an approach to correct corrupt labels. 
    
    \item We propose STRisk, a socio-technical predictive system that combines technical and Twitter signals to predict whether an organization will experience a hacking breach.
    
    \item We conduct a qualitative analysis of the most influential indicators to highlight those that need to be looked after, when dealing with hacking breaches.

\end{enumerate}

The rest of the paper is organized as follows: in section II we discuss the related literature. Section III gives details on the datasets used in this study and the performed data processing. Section IV explains the process of feature extraction. In section V We introduce our approach to handle noisy labels, and present our prediction methodology in Section VI. In Section VII we present and discuss the obtained results, and in Section VIII we discuss the limitations of this work and provide directions for future work. Finally, conclusions are drawn in Section IX.

\section{Literature Review}

Cyber threats prediction is an emerging research direction dealing with cybersecurity threats in a proactive fashion. Similarly, social media and Twitter signals are increasingly being used by the research community to study various topics. For example, Twitter was used to predict stock markets \cite{bollen2011twitter} and elections \cite{yaqub2017analysis} as well as to understand emerging cyber threats \cite{hernandez2018social}. As our work considers both technical and social aspects to predict data breaches, we then review previous work from three different perspectives: technical, social and socio-technical.

\subsection{Breach incidents prediction from technical viewpoint}

Most of data breach prediction studies focus on technical indicators to predict the occurrence likelihood of breach incidents. One the closest works to our study is the paper by Liu \textit{et al.} \cite{liu2015cloudy} who conducted an external assessment of an organization's security posture. They aggregated 258 measurable features such as network mismanagements (DNS misconfiguration, open recursive resolvers, untrusted certificates), and malicious activities (phishing and spam). They found that mismanagements have the highest importance value in the predictive model but are poor predictors against malicious activities. Similarly, Bilge \textit{et al.} \cite{bilge2017riskteller} presented RiskTeller to study 600K machines and predict their risk of infection by malware by building a profile consisting of 86 features for each machine. The profile includes volume of events, applications installed, patching behavior and threat history. They found that the machines that present a high risk of infection are characterized by few or no security upgrades and unusual usage in terms of temporal patterns and binaries installed. Additionally, Edwards \textit{et al.} \cite{edwards2019risky}, in an assessment of organizational risk to botnet infections, found that peer-to-peer file sharing correlates strongly with botnet infections, and organizations that experienced breaches had 67.8\% and 55.5\% of botnet and peer-to-peer file sharing activity respectively. In the same vein, in \cite{sarabi2016risky}, Sarabi \textit{et al.} leveraged technical features such as IP Intelligence and Alexa Web Information Service (AWIS), and organizational information from Open Directory Project (ODP) on organizations to assess the organization's risk of experiencing a data breach. Diversely, Xu \textit{et al.} \cite{xu2018modeling} approached the problem of hacking beaches prediction from a statistical perspective by looking at the intrinsic patterns the past data breaches exhibit including inter-arrival time of breaches and autocorrelations within the time series of the number of breaches and their sizes. On the PRC dataset, Xu \textit{et al.} \cite{xu2018modeling} found that breaches are getting worse in terms of frequency which does not align with the conclusions drawn by Edwards \textit{et al.} \cite{edwards2016hype} suggesting that neither frequency nor breach size has increased over the period between 2005 and 2015.

On the other hand, Sen and Borle \cite{sen2015estimating} used opportunity theory of crime to assess the data breach occurrence risk in a specific industry and the risk of occurrence of a specific breach type. They found that investing in IT security correlates with a higher risk of experiencing a data breach, and applying strict data breaches reporting laws could lead a reduced risk of data breach. Lower-reputation and higher-reputation firms in terms of stock market value was studied by Gwebu \textit{et al.} \cite{gwebu2018role} and found that firms with lower reputation suffer a significant negative return in stock value, in contrast to high reputation firms. This indicates that they are investing more in IT security, allowing them to recover quickly from a breach. In the same vein, Romanosky \cite{romanosky2016examining} found that costs related to data breaches exhibited an increasing curve starting from 2008, and suggests that firms lack strong motives to invest in data security and privacy protection. Besides, Romanosky \textit{et al.} \cite{romanosky2011data} studied whether the application of data breach disclosure laws reduces the the identity theft. They used an econometric model and found that disclosure laws reduce identity theft by 6.1\%. On a security management level, Ikegami and Kikuchi \cite{ikegami2020modeling} found that the inter-arrival time is 10\% longer when an external audit is conducted.

\subsection{Social media for cyber attacks understanding}

Social media and particularly Twitter has been used in the context of cyber attacks mainly to understand and detect threats being discussed and track disclosed vulnerabilities. For example, Khandpur \textit{et al.} \cite{khandpur2017crowdsourcing}, used Twitter for crowdsourcing cyber attacks, and managed to identify data breach related Twitter events with an accuracy of 71\%. In addition, Shu \textit{et al.} in \cite{shu2018understanding} used social media as a sensor to understand cyber attacks from social behaviors, and to quantify the impact of negative sentiments on the occurrence of attacks such as malicious emails and endpoint malware. Twitter was also used to tell if a vulnerability is exploited or not as presented in Sabottke \textit{et al.} work \cite{sabottke2015vulnerability}. Ritter \textit{et al.} \cite{ritter2015weakly} considered Twitter to detect the presence of cyber attacks such as denial of service, data breaches and account hijacking.

Sarkar \textit{et al.} \cite{sarkar2019predicting} studied darkweb dicussions to collect information on vulnerabilities being discussed mainly by the most influential users in order to generate warnings on future attacks. Horawalavithana \textit{et al.} \cite{horawalavithana2019mentions} used information on vulnerabilities from Twitter and Reddit to predict activities in Github repositories in response to the disclosed vulnerabilities as an early alert tool of public vulnerabilities discussed on social media.

\subsection{Cyber attacks from socio-technical perspective}

Combining both social and technical dimensions for prediction has not been widely addressed. Among the few studies in the literature, Okutan \textit{et al.} \cite{okutan2018forecasting} used unconventional signals from GDELT, Twitter and OTX to predict malicious email, malicious destination and endpoint malware threatening a single organization. Similarly, Goyal \textit{et al.} \cite{goyal2018discovering} used signals from darkweb, Twitter and security blogs to predict endpoint malware, malicious email and malicious destination within two organizations using time series forecasting. In another context, Twitter sentiments used as exogenous signals proved to improve the prediction performance of probing rates by 5\% \cite{hammouchi2019predicting}. Similarly, Almukaynizi \textit{et al.} \cite{almukaynizi2017predicting} applied social network analysis on darkweb forums to predict the exploitability of vulnerabilities, and showed that social features improve the f1-score by about 6\%.

\subsection{Synthesis}

Overall, the aforementioned studies have highlighted important aspects that have demonstrated good performance in risk assessment, such as technical mismanagements within organizations' networks. However, most of the previous work has approached data breaches prediction mainly from a technical perspective and has neglected other interfering aspects such as the social dimension. Moreover, to the best of our knowledge, no work has addressed the problem of unreported incidents and has only considered reported data breach incidents. In this paper, we address data breaches caused by hacking activities by expanding the scope of prediction and considering both security and social postures. In doing so, we uncover the hidden social aspect of data breaches and provide organizations with a holistic view of their risk of exposure to hacking breaches. We also propose an approach to deal with noise in the non-victim sample and attempt to detect unreported incidents. Our approach formalizes technical security posture as network misconfigurations and malicious activities, and social posture as signals from Twitter. 

\section{Data Collection and Processing}

In this section, we present the data collection process where we build sets of victim and non-victim organizations associated with their technical security posture and social reputation measured externally. Our data collection covers the period from January 2016 to September 2019. 

\subsection{Breach Incidents Records}

Building a ground truth of breach incidents starts from collecting the past incidents from reliable sources. To this end, we collect incidents reported in two sources, namely privacy rights clearinghouse dataset, and veris community database. 

\subsubsection{Privacy Rights Clearinghouse Data}
Privacy Rights Clearinghouse (PRC) \cite{PRC} is a non-profit organization that collects and reports data breaches recorded mainly in the USA since 2005. Each data breach incident comes with a short description, known date of breach, number of breached records, and type of breach. Each breach can be caused by an insider that intentionally breaches information (INSD), a payment card fraud (CARD), a physical loss (PHYS), a lost or stolen portable device (PORT), being hacked by someone or infected by malware (HACK), a stationary equipment loss (STAT), an unintended disclosure (DISC) and an unknown method (UNKN). In our study we retain only breaches caused by hacking activities (HACK).

\subsubsection{Veris Community Database}
Veris Community Database (VCDB) \cite{Veris} is a project maintained by Verizon RISK Team and released in 2013 to collect and disseminate security incidents information from different reliable sources. The current version contains more than 8000 incidents and are divided into 7 categories called threat actions: malware, hacking, social, misuse, physical, error, and environmental. In the present study, we focus on malware and hacking as they are directly linked to hacking activities by outsiders.

In order to align the incidents with the other datasets, we extract incidents occured between Jan-2016 to Sept-2019 reported in PRC and VCDB. After filtering out redundant and non-hacking incidents, we keep a total of 795 unique incidents. By combining incidents from PRC and VCDB we reduce the bias of missing reported incidents among the publicly disclosed records. Table \ref{incident_example} shows a snapshot of reported incidents in both PRC (2) and VCDB (1).

\begin{table}[h]
    \centering
    \caption{An example of a hacking incident record caused by exploiting a misconfiguration reported in VCDB (1) and a another hacking breach discovered unintentionally via an internet scan as reported in PRC (2)}
    \label{incident_example}
    \begin{tabular}{|c|l|p{4.5cm}|}
    \hline
    & Attribute & Value \\ \hline
    \multirow{6}{*}{(1)} & Victim Organization & Librer Porr \\
    & Asset data amount &  2,100,000  \\
     & Hacking notes  &  Direct server access from internet \\
    & Asset notes	 &   mongo-DB \\
    & Exploiting misconfig.  & TRUE  \\
    & Incident summary  & { \raggedright \textit{Breach 2 for github ID. Database deleted and randsomed }}  \\
     \hline \hline 
    \multirow{3}{*}{(2)} & Victim Organization &  Schoolzilla \\
    & Breached records  & 1,300,000	\\
    & Incident summary &  
    { \raggedright
\textit{More than a million American students had their information exposed including social security numbers. A researcher discovered the Schoolzilla breach while scanning the web for an “all too common” misconfiguration in amazon cloud storage devices}}
     \\ \hline
     
    \end{tabular}
\end{table}

\subsection{Non-Victim Records}

In order to build a dataset consisting of both victim and non-victim organizations for the predictive models, we add a non-victim sample to our ground truth. In addition to the victim organizations presented above, we randomly select a sample of non-victim organizations (negative sample) from the American Registry for Internet Numbers (ARIN) with a size {$\sim(\times4)$} larger than the victim set (positive sample). The choice of the negative sample size is motivated by the common belief that the number of attacked organizations is smaller than that of attacked organizations. We choose our non-victim sample to be disjoint from to the victim sample, i.e.:

\begin{equation*}
    \{{\rm Victim\ Set}\} \cap \{{\rm Non\text{-} Victim\ Set}\} = \emptyset
\end{equation*}

It is noteworthy that there may be some organizations in the non-victim set that could have experienced a breach in the past but has not been reported. We attempt to solve this issue by proposing a noise correction approach in section \ref{data_labeling}.
\subsection{Technical and Misconfiguration Data}

Considerable amount of information can be gathered from the outside of an organization. This allows to gain insights into the security posture of organizations without knowing their internal structure and policy. To do so, we first map organizations to their IP ranges using the internet registry, and refer to scanning and spamming activities, IP blacklists, and web certificates to quantify the technical security posture. 

\subsubsection{ARIN Registry}

An internet registry or regional internet registry (RIR) contains network IP-level network information about organizations. A RIR keeps track of IP allocations and registrations to organizations and individuals in different regions of the world. There exist five regional internet registries (RIRs): ARIN \cite{ARIN} for America, AFRINIC \cite{AFRINIC} for Africa, APNIC \cite{Apnic} for Asia-Pacific, LACNIC \cite{Lacnic} for Latin America, and RIPE NCC \cite{RipeNCC} for Europe, Central Asia, Russia, and West Asia. 
Since our study focuses on data breaches in the United States of America, we use the ARIN registry to identify the IP ranges of selected organizations.


In order to map the organizations to their respective IP range, we use the ARIN registry to retrieve the IP addresses. This task is not straightforward as the names of organizations in the incidents datasets generally differ slightly from the names registered in ARIN as shown in table \ref{tab_matching_names}. In order to find the corresponding names, we refer to similarity metrics for strings. In our case, we perform the matching in two steps using two similarity metrics, namely Jaccard similarity and Jaro-Winkler similarity defined as follows: 
\begin{equation}
    Jaccard(N_i, N_r) = \frac{|N_i \cap N_r|}{|N_i \cup N_r|} 
    \label{jaccard_eq}
\end{equation}
\begin{equation}
    JW(N_i, N_r) = Jaro(N_i, N_r)+ \ell_1 p(1-Jaro(N_i, N_r))
    \label{jaro_wink}
\end{equation}
where \begin{equation}
    Jaro(N_i, N_r) = \bigg\{ \begin{array}{ll}
        0 &  {\rm if} \, m=0\\
        \frac{1}{3}(\frac{m}{|N_i|}+\frac{m}{|N_r|+\frac{m-t}{m}} & {\rm otherwise}
    \end{array}
    \label{jaro}
\end{equation}
with $N_i$ being the name in the incident dataset, $N_r$ being the name in the ARIN registry, $m$ being the number of matched characters in the two names, $t$ being half of the number of transpositions, {$\ell_1$} being the length of common prefix at the start of the string, and $p$ being a scaling factor ($=0.1$).

We first use the Jaccard similarity to refine the names, as this similarity examines the number of words shared between two sets (names). Second, we use the Jaro-Winkler (JW) similarity to refine the match between the two samples, as this similarity looks at the longer prefix corresponding to the beginning of two names. Both similarity measures are useful in our case as we have names with suffixes such as 'LLC', 'DBA', 'INC' ...\}, or names that we need to match by looking at the beginning of the string. After the refinement using the similarities, we manually check the match. For example in table \ref{tab_matching_names} `florida bar association' and `florida bankers association' are two different organizations and this mismatching error is successfully detected when using Jaccard similarity. On the other hand, the mismatching between `white house' and `whitehouse hotel' is not detected using similarity metrics, and is thus fixed manually.

After retrieving the matched names, we check them manually to verify the integrity of the matching, then we map the corresponding IP ranges of the full set of organizations.
\begin{table}[h]
    \centering
    \caption{Examples of matchings and mismatchings detected and undetected by similarity metrics}
    \label{tab_matching_names}
    \begin{tabular}{|c|c|c|c|}
    \hline
   Real name & Matched name & Method & Matching error \\ \hline
\makecell{florida bar \\ association} & \makecell{florida bankers \\ association} & Jaccard &  Detected \\
    nsc technologies &  ssc technologies &  Jaro-Winkler & Detected \\ 
    white house & whitehouse hotel & & Not Detected \\
    trump hotels &  triumph hotels & & Not Detected \\
    \hline
    \end{tabular}
\end{table}

\subsubsection{TCP Scans}
Many organizations perform regular internet wide scans to identify open ports on hosts, certificates, and many other information on a given IP. In this work, we use Rapid7 \cite{rapid7} scan TCP data, maintained within its Sonar project. For each IP address we look at the open ports over the period of study. This project on listing open ports and services across all the IPv4 space has started in Mar-2017. We collect all shared information since then. 

\subsubsection{FireHOL BlackLists}
Blacklists provide basic security against potential attackers by listing the malicious domains and IPs that should be blocked. Firehol blacklists \cite{FireHOL} aggregates a set of well known and reliable blacklists such as abuse.ch, dshield, spamhaus, team-cymru, and many others which give trusted information on malicious IPs and domains. For each organization we look for the blacklisted IPs which may be related to or involved in malicious activities and may lead to breaches.

\subsubsection{Darknet}
Darknet or network telescope is a passive monitoring system with no services running behind. It logs every attempt of establishing a connection. For each connection, several information are recorded such as IP source, IP destination, protocol number, TCP flag, TCP SYN, TCP source port, TCP destination port, \textit{etc}. We consider the darknet as another blacklist since the allocated IP addresses are not registered (not known to the public), and if an IP belonging to a given organization is observed on the darknet, then this host may be compromised and used by attackers (e.g. as a rebound) to participate in attack campaigns. In this work, we use a /24 darknet (4096 allocated IP addresses) hosted at Inria in France.

\subsubsection{Spamming Activities}
To track blacklisted spam domains we use JwSpamSpy \cite{jwSpamSpy}, a daily updated list of blacklisted spam domains which is maintained by joewein.de LLC, a software company specialized in tracking spam and online fraud. To lookup blacklisted domains in our dataset, we first map each organization's IP with its name server (ns) by performing a reverse dns lookup (RDNS). For this task, we use Rapid7 daily RDNS records to get the dns associated with each IP address. Next, we use the $ns$ nameservers of each organization to search for spam domains.

\subsubsection{SSL Certificates}

To analyze SSL certificates (X.509 certificate) on each endpoint of a given organization, we use the sonar SSL dataset maintained by Rapid7 \cite{rapid7} which consist of metadata observed when communicating with HTTPS endpoints over the entire IPv4 space. Each scan record consist of four files: \textit{hosts} and \textit{endpoints} that map between IPs/endpoints and the certificate fingerprint, \textit{certs} map the certificate with its corresponding fingerprint, and the file names that maps between the certificate \textit{name} and its fingerprint. We join and merge all files using their hashes and check if a certificate associated with an IP is valid or has expired. 

\subsection{Social Data}
Social media, particularly Twitter, have been used as a source of intelligence and public tone measurement in many applications ranging from stock markets, elections, to cyber-attacks. Twitter is a reliable source to acquire information on recent vulnerabilities and cyber-attacks attempts. In the present study, we use Twitter to gain insights into US organizations' social reputation and design a set of features that characterize this reputation.

\subsubsection{Sector and Twitter handle lookup}

In order to get social data about an organization, we proceed by the name of organizations and/or their Twitter handles. For the sector of organizations we refer to \textit{Crunchbase} platform \cite{crunch}, which provides information about private and public businesses. First, we get the corresponding \textit{url} on crunchbase for each organization by querying on Google "\textit{{crunchbase: organization name}}" using Google-search python API. We use the retrieved links to scrape crunchbase and parse information consisting of mainly the sector, number of employees, links to social media accounts such as Twitter, facebook and linkedin. For our study, we retrieve the sector and Twitter url. The sectors are categorized into 10 main sectors as detailed in table \ref{sect_distr_tab} with the portion of victim and non-victim organizations within each sector.

\begin{table}[h]
    \centering
    \caption{Distribution of incidents by sector with their corresponding percentage in the whole dataset, median size of organizations by sector, and victim and non-victim distributions }
    \label{sect_distr_tab}
    \begin{tabular}{|l|c|c|c|c|}
    \hline
    Sector & Percent & \makecell{Median \\ Org. size} & Non-victim (\%) & Victim (\%) \\ \hline
    \makecell[l]{Information- \\ \,Technology} & 33.5\%  & 32 & 90\% & 10\% \\
    \makecell[l]{Medical- \\ \, \& Healthcare}   & 14.5\%  & 16 & 50\% & 50\% \\
     Finance & 11.5\%  & 32 &  77\% & 23\% \\
     Retail & 9.5\%  & 32 & 79\% & 21\% \\
     Education & 9\%  & 96 & 52\% &  48\% \\
     Entertainment & 8\%  & 32 & 88\% &  12\% \\
     Industrial  & 6.5\%  & 32 & 52\% &  48\%   \\
     Government & 4.5\%  & 40 & 65\% &  35\% \\
     NGO & 1.5\%  & 32 & 72.5\% &  27.5\% \\
     Energy & 0.5\%  & 32 & 97\% &  3\% \\ \hline
    \end{tabular}
\end{table}

\subsection{Twitter scraping}

Collecting tweets about organizations is a bit challenging since we do not have access to the premium Twitter API and the free version does not allow to go back in time for more than one week. Thus, we refer to a scrapping tool to crawl Twitter and search by organizations' names or their Twitter handles. For that, we use Twitter urls scraped from crunchbase and use twitterscraper tool \cite{twitterscrapper} to scrape tweets. We scrape all English tweets from Jan-2016 to Sept-2019 using organization's twitter handle if available or organization's name as a query. We collect a total of 2.5M tweets. Twitterscraper returns several information and statistics on tweets including: hashtags, is replied to (if the tweet is replied to), is reply to (if the tweet is a reply to another tweet), usernames, likes count, tweet text, retweets count, handles, and replies.

\begin{table}[h]
    \centering
    \caption{Examples of Tweets collected on victim and non-victim organizations}
    \label{tab:my_label}
    \newcommand\T{\rule{0pt}{2ex}}       
    \begin{tabular}{|l|p{4.5cm}|c|}
    \hline
    Company & Example Tweet & Class \\  \hline 
    \hline
        CPanel & Hackers planted a backdoor in \#Webmin-a popular cPanel type utility for \#Linux servers—that remained hidden for over a year, allowing unauthenticated remote attackers to execute commands on affected servers with root privileges & \makecell*{Victim \\(Breach)}  \\  \hline
        \makecell*[cl]{LookingGlass \\Cyber \\Solutions}  & \vspace{-0.4cm} Voting records of 40 million + Americans for sale on dark web \text{@}LG\_Cyber \#cybersecurity &  \makecell*{Victim \\(Breach)}   \\ \hline
        \makecell*[ml]{GE\\ Healthcare} & In medicine you need a randomized controlled trial to know whether a new diagnostic test is working or not. Anything less is just fake science and fake medicine. & \makecell*[l]{Non-victim \\(No-Breach)}   \\ \hline
        \makecell*[l]{Bloom \\Energy} & 451 Research outlines our plans to leverage Bloom Energy Fuel Cells to create more efficient data centers. & \makecell*{Non-victim \\(No-Breach)} \\ \hline
    \end{tabular}
    
\end{table}

\section{Feature Extraction}

After collecting technical and social data, we proceed by extracting and measuring the indicators that we use to assess the security and social postures of organizations and predict their risk of exposure. 

\subsection{Misconfiguration lookup}

In order to search for misconfigurations and malicious activities for each organization, we use IP ranges and domain name of each organizations. For open ports, blacklists, darknet, SSL certificates we proceed by IP. For each organization's IP range we check whether it is listed in malicious activities and have network misconfigurations. For each organization we extract the no. of hosts (IP) having misconfigurations (\# of open ports, \# of blacklisted IP addresses, \# of expired certificates) and the ratio (No. of misconfigurations / No. of hosts). As for the domain name, we use it to search for domains involved in spamming and count the number of spam domains as well as calculating the ratio.

\subsection{Social signals}

From the collected raw tweets, we extract the following features for each organization:
        
\begin{enumerate}[label=(\alph*)] 
    \item No. of tweets mentioning the organization (mentions) 
    \item No. of unique accounts tweeting about the organization (popularity)
    \item No. of retweets / No. of tweets about the organization (spreadability)  
    \item No. of retweets / No. of replies (debatability: if the tweet has more replies than retweets, it means that people object on the content)
    \item No. of likes / No. of tweets about the organization (agreeability)
    \item Average (avg.) likes
\end{enumerate}
In order to get tweets sentiments, we first clean the tweet text from emojis, special characters and links. Next, we replace reduced words such as \{he/she's, they're, they've \} with the complete words \{he/she is, they are, they have \} to avoid any perturbation to the sentiment polarity measurement. After cleaning the text, we use the TextBlob tool which is a Natural Language Processing tool built upon Stanford NLTK tool, to get the polarity of a tweet. Given the polarity ($p$), we distinguish between five different sentiments as suggested in  \cite{kausar2019sentiment}. In our categorization, we consider as neutral every tweet having a polarity between -0.1 and +0.1. Overall, sentiments classification is given as follows:

\begin{tabular}{@{}lc}
\\ \normalsize
    Strong negative & $-1<=p<-0.5$ \\
    Weak negative  & $-0.5<=p<-0.1$ \\
    Neutral & $-0.1<=p<=-0.1$ \\
    Weak positive  & $0.1<=p<=0.5$ \\
    Strong positive & $0.5<p<=1$ \\ \\
\end{tabular}

Altogether, in table \ref{feature_set__} we summarize technical and social feature sets extracted from the aforementioned sources. Additionally, Table \ref{dataset_feats_stats} provides details about the distributions of different features used in this study

\begin{table}[h]
    \centering
    \caption{Summary of technical and social feature sets used in this study}
    \label{feature_set__}
    \begin{tabular}{|l|p{7cm}|}
    \hline
        \makecell[l]{Technical \\ features} & \# of blacklisted IPs, blacklised IPs ratio, \# of IPs on darknet, ratio of IPs on darknet, \# of open ports, open ports ratio, \# of expired certificates, expired certificates ratio, \# of spam domains, spam domains ratio  \\ \hline
        \makecell[l]{Social \\ features} &  \# of Non-organization tweets, \# of unique accounts,  \# of Non-organization retweets, \# of Non-organization replies, \{\# of retweets / \# of tweets\}, \{\# of retweets / \# of replies\}, \{\# of replies / \# of tweets\}, \{\# of is\_reply\_to / total \# of tweets\}, \{\# of replied\_to / \# of tweets\}, \{\# of likes / \# of tweets\} (avg. likes), \# of strong negative sentiments, \# of weak negative sentiments, \# of neutral sentiments, \# of weak positive sentiments, \# of strong positive sentiments, avg. polarity 
        \\
        \hline
    \end{tabular}
\end{table}

\begin{figure*}[h]
    \centering
    \includegraphics[width=\linewidth]{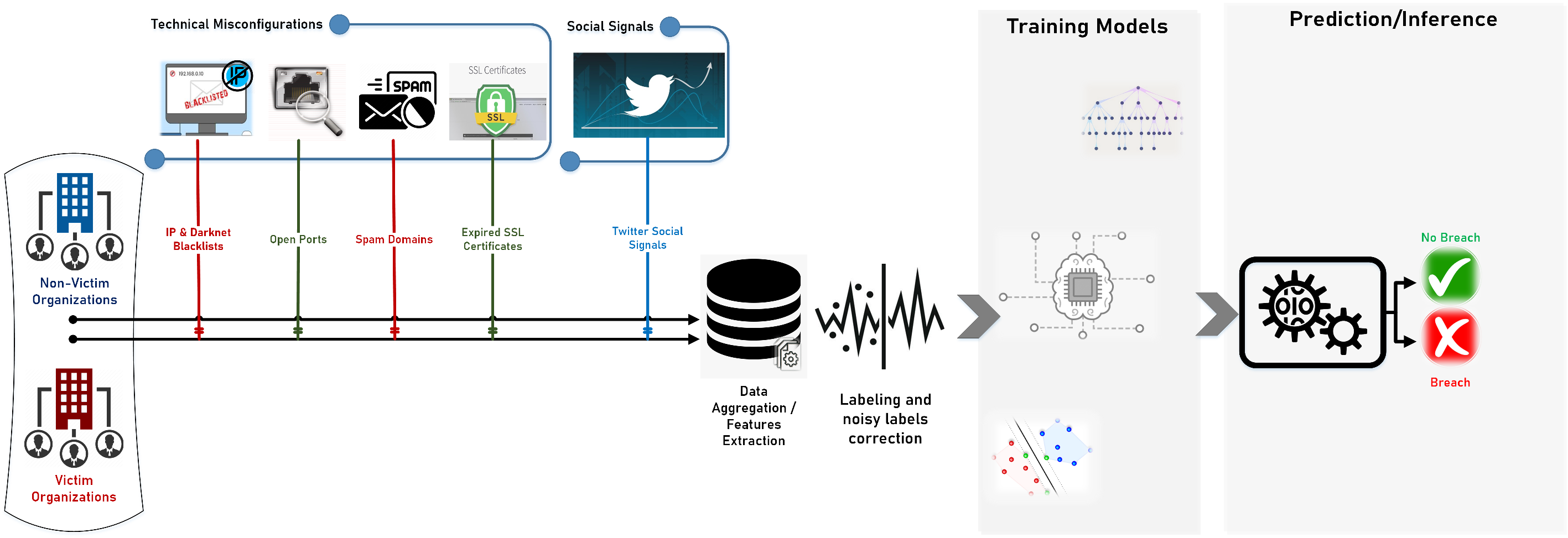}
    \caption{STRisk pipeline to combine technical misconfigurations and Twitter social signals for both victim and non-victim organizations, correct noisy labels and build the predictive models to discriminate risky organizations from non-risky ones}
    \label{model_pipeline}
\end{figure*}

\begin{table*}[h]
    \centering
    
    \caption{Statistics of Socio-technical Features for both victim and non-victim organizations}
    \label{dataset_feats_stats}
    \setlength{\tabcolsep}{2pt}
    \resizebox{\textwidth}{!}{
    
    \begin{tabular}{lrrrrrrrrrrrrrrrrrrrr}
    \hline
    {} & Org Size &  \makecell{Blacklist \\Ratio} &  \makecell{Darknet\\ Ratio} & \makecell{Expired \\Certificates\\ Ratio} &  \makecell{Open\\Ports\\ Ratio} &  \makecell{Spam \\Domains\\Ratio} &  \makecell{Unique \\ accounts} &  \makecell{Weak \\ neg.}  &  \makecell{Strong \\ neg.} &   \makecell{Weak \\ pos.} &  \makecell{Strong \\ pos.} &     Neutral &  Mentions & Replies &  \makecell{Retweet \\ ratio} & Debatability & \makecell{Avg. \\ likes}  &  \makecell{Avg. \\ polarity}  &  \makecell{Likes \\ ratio}  & \makecell{Reply \\ ratio } \\
    \hline
    
    \hline
    Mean  &    1.600 &         1.09 &       0.081 &                    0.73&          9.37 &            0.23 &       309.38 &      46.95 &         5.68 &     209.68 &        48.23 &   343.64 &   654.19 &     165.85 &        0.93 &      5.05 &     2.52 &      0.13 &     2.86 &       0.149 \\
    STD   &   414912 &         1.80 &       0.105 &  1.18 &  33.32 &  0.19 &       289.94 &      52.19 &  9.20 & 176.77 &  52.87 &   281.99 &   492.71 &     241.87 &       11.43 &     10.65 &    31.27 &      0.092 &    31.30 &       0.34 \\
    Min   &       1 &             0.00 &           0.00 &                        0.00 &              0.00 &                0.00 &             0.00 &           0.00 &             0.00 &           0.00 &             0.00 &     0.00 &      0.00 &           0.00 &           0.00 &          0.00 &       0.00 &         0.00 &        0.00 &          0.00 \\
    
    Q1   &      8 &         0.00 &       0.00 &                    0.00 &          0.00 &            0.00 &        35.0 &       4.00 &         0.00 &      38.00 &         5.00 &    59.00 &   119.00 &       8.00 &        0.20 &      1.00 &     0.40 &      0.072 &     0.61 &       0.03\\
    
    Median   &      32 &         3.47 &       1.01 &                    0.01 &          5.51 &            0.56 &       221.00 &      31.00 &         2.00 &     205 &        36.00 &   327.00 &   713.00 &      73.00 &        0.486 &      2.96 &     1.166 &      0.123 &     1.507 &       0.126 \\
    
    Q3   &    220  &         11.03 &       2.31 &                    0.22 &          19.84 &            1.85 &       549 &      77.00 &         8.00 &     338.00 &        76.00 &   570.50 &  1122 &     262 &        0.92 &      5.83 &     2.37 &      0.176 &     2.924 &       0.284 \\
    Max   &  22404 &        98 &       7.25 &                   18.25 &        660.75 &            5.32 &       971 &     566 &       118 &    1381 &       622 &  1336 &  1495 &    5511 &      698.14 &    324.01 &  1903.47 &      1 &  1903.47 &       4.68 \\
    \hline
    \end{tabular}} 
    
\end{table*}



\section{Data Labeling}
\label{data_labeling}
In order to distinguish between victim and non-victim organizations using supervised predictive models, we need to assign labels to each organization. In our case, we distinguish between victim organizations (class 1) and non-victim organizations (class 0). As simple as this task may seem, the information we have about breach incidents is only valid if all victim organizations have reported being victims of a breach. On this basis, we are certain that the reported incidents are true and should be assigned $\textbf{1}$ (class 1), but we are not certain that the organizations in the negative sample are truly non-victim, as they may have experienced a breach and did not report it or missed it. This problem is known as learning from noisy labels, and we consider correcting the noise in the negative sample before predicting the exposure of organizations. In the following, we present our approach to solve this problem and clean our data for reliable risk assessment and prediction.

In this context, several approaches have been proposed in the literature to learn and build robust models in the presence noisy labels. In this context, loss reweighting was proposed to take into account the noise rate and inject it into the objective function \cite{natarajan2013learning}. However, loss reweighting methods imply knowing noise rate in each label which is not always trivial. Learning from Positive-Unlabeled data known as PU learning is another technique to learn from positive and unlabeled data \cite{elkan2008learning}. This method treats unlabeled samples as weighted negative samples to identify corrupt labels. Another technique is learning with confident examples known as confident learning \cite{northcutt2019confident}. Confident learning relies on ranking and pruning noisy examples to train with confident ones. In our case, we are interested in finding noisy labels and flip their class value. Thus, we focus only on the negative sample since the positive sample is perfectly labeled. 

\subsection{Naive Labeling}

The first approach to labeling the data at hand is to assign a label of 1 to organizations featured in PRC and VCDB datasets and temporarily label the non-victim organizations as Class 0. This approach is valid if we are certain that the organizations in the negative sample did not experience a breach, which is not the case.

\subsection{Noisy Labels Discovery}

Based on the naive labeling, we attempt to find corrupted labels in the negative sample and correct them. We assume that labels of the negative sample are noisy and every example may be mislabeled as class 0 when it should belong to class 1. Let $X:=(x,\widetilde{y})^n \in (\mathbb{R}^d,{0,1})^n$ be our dataset of $n$ examples associated with noisy labels $\widetilde{y}$. In our case we assume that the noisy labels $\widetilde{y}$ are only labels of negative sample (class 0) and we denote latent true labels as $y^*$. We assume that features $x_i$ are noise free and we consider handling only the noise in the labels.  

In order to find the true labels of examples in the negative sample, we estimate the two probabilities $p(\widetilde{y}|y^*)$, the probability of the noisy label $\widetilde{y}$ given the true label $y^*$ and $p(y^*)$ the probability of the true label. Inspired by Confident Learning \cite{northcutt2019confident}, we estimate $p(\widetilde{y}|y^*)$ and $p(y^*)$ jointly as $p(\widetilde{y},y^*)$ using confusion matrix and confident joint defined as:
\begin{equation}
    C_{\widetilde{y}, y^*} [i=0][j=1] := |\hat{X}_{\widetilde{y}=0, y^*=1}|
\end{equation}

where $\hat{X}_{\widetilde{y}=0, y^*=1} := \{ x \in X_{\widetilde{y}=0}:\hat{p}(\widetilde{y}=1);x,\theta) \geq t_{j=1} \}$

where $\theta$ is any probabilistic model, $\hat{p}()$ the out-of-sample prediction probability given by the model $\theta$, and the threshold $t_{j=1}$ is the expected self confidence for class $j=1$ defined as: 

\begin{equation}
    t_{j=1} = \frac{1}{|X_{\widetilde{y}=1}|} \sum_{x \in X_{\widetilde{y}=1} } \hat{p}(\widetilde{y}_{j=1}; x,\theta)
\end{equation}

To estimate the noise transition matrix and find mislabeled examples, we consider the confident joint $C_{\widetilde{y}, y^*}$ with the threshold $t_{j=1}$ and the confusion matrix $CM$ with the threshold $0.5$. Given the confident joint and confusion matrix, we estimate the noise transition matrix by normalizing rows and dividing by total out-of-sample examples as follows:

\begin{equation}
    \hat{Q}_{\widetilde{y}=i,y^*=j} = \frac{\frac{C_{\widetilde{y}=i,y^*=j}}{\sum_{j \in \{0,1\}} C_{\widetilde{y}=i,y^*=j}} \cdot |X_{\widetilde{y}=i}| } { \sum_{j \in {0,1}} \left(  \frac{C_{\widetilde{y}=i,y^*=j}}{\sum_{j \in \{0,1\}} C_{\widetilde{y}=i,y^*=j}} \cdot |X_{\widetilde{y}=i}| \right) }
\end{equation}

The noise transition matrix in our setting can be easily obtained based either on the confident joint and confusion matrix by dividing by the sum over rows of the matrix. For simplicity, we denote both confident joint $C_{\widetilde{y}, y^*}$ and confusion matrix $CM_{i,j}$ by $Z_{i,j}$. So the transition matrix based on $Z_{i,j}$ (confident joint or confusion matrix) is as follows:

\begin{equation}
    \hat{Q}_{\widetilde{y}=i,y^*=j} = \frac{Z_{i,j}}{\sum_{i^\prime \in \{0,1\}} Z_{i^\prime ,j}}
\end{equation}

From the matrix $Z_{i,j}$ in table \ref{z_ij_matrix}, in our case we focus on estimating the noise probability $0 \rightarrow 1$ (i.e. false negatives)

\begin{table}[H]
    \centering
    \caption{$Z_{i,j}$ Matrix representing confident joint when threshold is $t_j$ and regular confusion matrix when threshold is $0.5$ }
    \label{z_ij_matrix}
    
    \begin{tabular}{|l|l|c|c|}
    \hline
    
       \multicolumn{2}{c|}{\multirow{2}{*}{\cellcolor{gray!25}{}}}  & \multicolumn{2}{c|}{\textbf{Predicted Outcome}} \\ \cline{3-4}
       \multicolumn{2}{c|}{\multirow{2}{*}{\cellcolor{gray!25}{}}}  & Breach(1) & No-Breach(0) \\ \hline \hline
       \multirow{2}{*}{{\textbf{Actual}} } & Breach (1) & \rule{2pt}{0pt} True Positive & False Positive\\ \cline{2-4}
     & No-Breach (0) & False Negative & True Negative \\ \hline
    \end{tabular}
    
\end{table}

\textbf{Noise discovery performance:} In order to test the validity of finding noisy labels using confident joint or confusion matrix, we flip randomly labels of a subsample from the positive sample ($N_{pos} = |X_{\widetilde{y}=1}|$) and use confident joint and confusion matrix to find corrupted labels. We repeat the experiment 10 times in cross-validation fashion by flipping randomly a subsample of size $(\frac{N_{pos}}{10})$ which is equivalent to injecting $10\%$ noise in the positive sample. To get the out-of-sample probabilities we use a bunch of models including Support Vector Machine (SVM), catBoost (CatB), lightGBM (LGB), logistic regression (LR), Bagging classifier (Bagg), and XGBoost (XGB). Then, we search for noisy labels using both confident joint and confusion matrix. We measure the average detection accuracy for both methods by calculating the ratio of right guesses. This means counting the number of examples in the flipped subsample predicted to be "class 1" over the subsample size $(\frac{N_{pos}}{10})$.

\begin{table}[h]
    \centering
    \caption{Mean noise detection accuracy using various models}
    \label{noise_det_acc}
    
    \begin{tabular}{|l|c|c|}
    \hline
        \multirow{2}{*}{Model} & \multicolumn{2}{c|}{Mean Noise Detection Accuracy}  \\ \cline{2-3}
        & Confident Joint & Confusion Matrix \\ \hline
        SVM &  25\% &  12.40\%  \\ \hline
        CatB &  64\% &  71.13\% \\ \hline
        LGB &  64.81\% &  71.39\% \\ \hline
        XGB &   \textbf{65.3\%}  & \textbf{74.81\% } \\ \hline 
        Bagg &    58.22\%   &    61.64 \%  \\ \hline 
        LR &  12\% &  11.89\% \\ \hline \hline
        SVM $\&$ CatB &  66.32\% &  71.64\% \\ \hline
        CatB $\&$ LGB &  69\% &  76.32\% \\ \hline
        CatB $\&$ Bagg &  68.1\% &  74.43\% \\ \hline
        XGB $\&$ Bagg &  68.22\% &  79.96\% \\ \hline
        CatB $\&$ LGB $\&$ XGB &  73.16\% &  81\% \\ \hline \hline
        All combined &  \textbf{76.58\%} &  \textbf{82.1\%} \\ \hline
        
    \end{tabular}
    
\end{table}

\subsection{Labels Correction}

\begin{figure}[h]
    \centering
    \includegraphics[width=\linewidth]{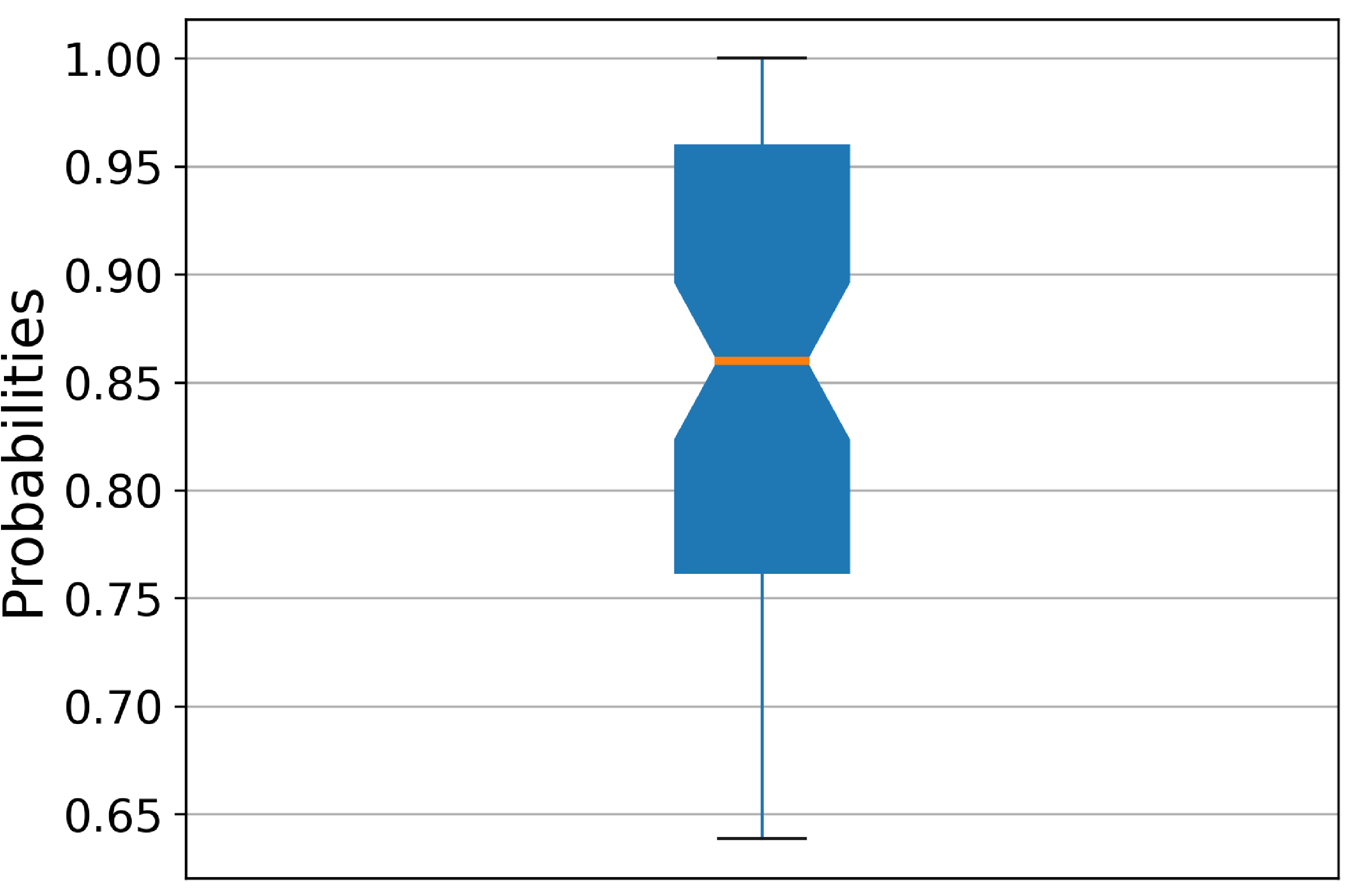}
    \caption{Boxplot of predicted probabilities for the examples that were selected to be flipped}
    \label{confident_boxplot} 
\end{figure}

As we are interested in detecting mislabeled organizations and to assess how well the socio-technical posture will predict the exposure of an organization to a data breach, we flip the class of detected corrupted labels from "class 0" to "class 1" instead of pruning them \cite{northcutt2019confident}. Based on results in \ref{noise_det_acc}, the best noise detection performance is achieved by combining the six models and using confusion matrix to estimate the noise. We use the six models combined to discover the corrupted labels in the negative sample, and flip their value to "class 1". For the examples detected as noisy, combining all models they predict that they are victim examples with high confidence. their probabilities range around 0.86 as shown in Fig. \ref{confident_boxplot}. The resulting distribution of labels after flipping is shown in table \ref{noise_corr}. In addition, Table \ref{noise_matrix} shows the conditional probabilities $ Pr(\widetilde{y} | \; y^*)$ of noisy labels given the latent true labels. The table shows $11\%$ of noise in the dataset, and $11\%$ of negative examples are corrupted and their labels should be flipped.

\begin{table}[h]
    \centering
    \caption{Labels distribution before and after noise correction}
    \label{noise_corr}
    \begin{tabular}{|c|c|c|}
    \hline
         & Breach (1) & No-Breach (0)  \\ \hline
        Before Noise Correction & 795  &  3016 \\ \hline
        After Noise Correction  & 1142 (+347)  &  2669 (-347)   \\ \hline
    \end{tabular}
    
\end{table}

\begin{table}[h!]
    \centering
    
    \caption{Noise matrix expressed in conditional probability of noisy label given true latent label }
    \label{noise_matrix}
    \begin{tabular}{|c|c|c|}
    \hline
        $ Pr(\widetilde{y} | \; y^*)$ & $ y^* = 0 $ & $ y^* = 1 $ \\ \hline
        $ \widetilde{y} = 0 $  & $ 0.89 $  &  $ 0.11 $ \\ \hline
        $ \widetilde{y} = 1 $ & $ 0.00 $ &  $ 1.00 $  \\ \hline
    \end{tabular}
    
\end{table}

\vspace{-5mm}
\section{Prediction Methodology}

\subsection{Problem Formulation}
\label{problem_formulation}
Our objective in this work is to assess hacking breaches risk and predict whether an organization will experience a hacking breach or not. We design the problem as a binary classification problem as depicted in figure \ref{model_pipeline} leveraging technical measurements and social signals detailed in table \ref{feature_set}. Each socio-technical profile is a vector of features consisting of $n$ measurements $\{X_i \in \mathbb{R}^n \:\ i\in \{1\cdots m \}\} $ representing the security and social postures of an organization. For each organization $i$ a predictive model produces a probability $P_i = P(infection\ |\ X_i)$ assessing the risk of experiencing a hacking breach. The produced risk probabilities are rounded to $0$ which means the organization is not at risk, and to $1$ which corresponds to a victim organization. 

We hypothesize that by approaching hacking incidents from a socio-technical perspective, we could better distinguish risky from non-risky (clean) organizations. Indeed, combining two complementary sources of information would enrich our understanding of at-risk organizations and highlight neglected but important aspects.

\subsection{Training-Inference Process}

\begin{figure}[h]
    \centering
    \includegraphics[width=\linewidth]{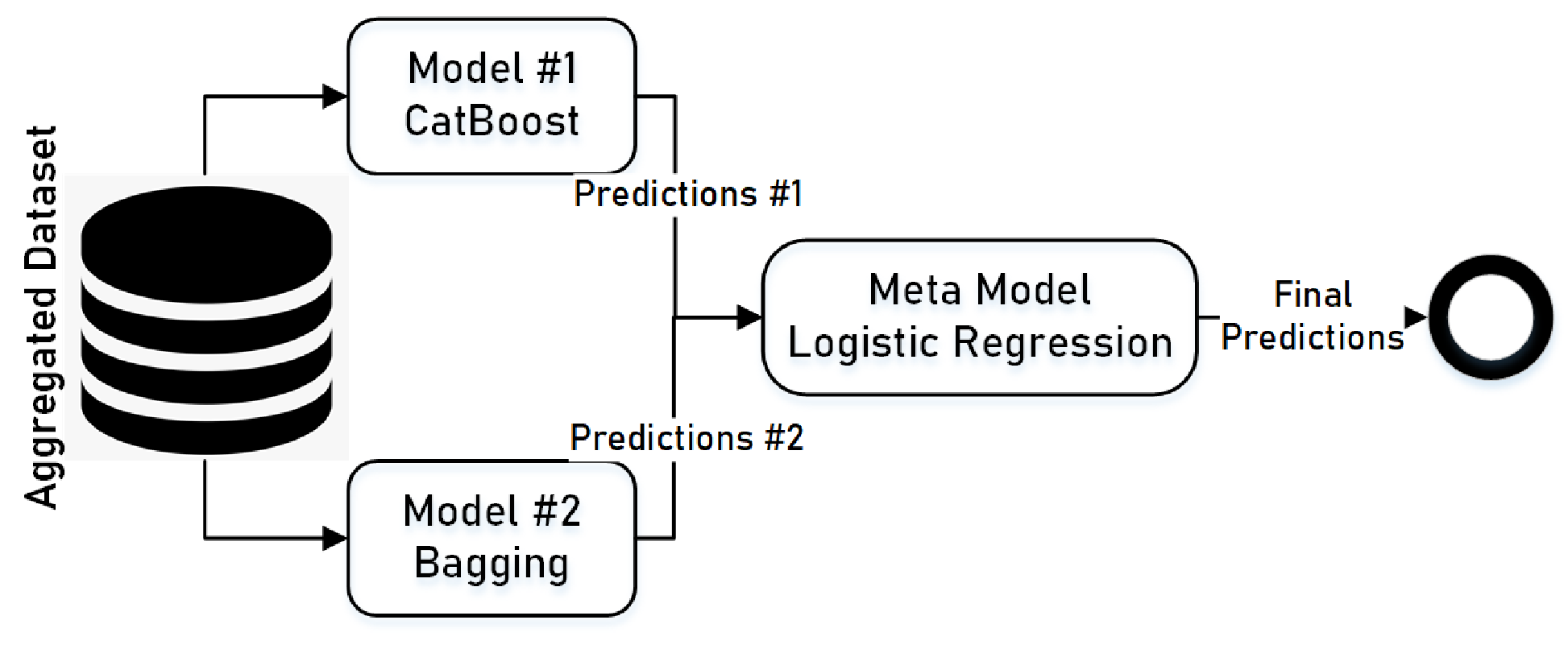}
    \caption{Example of stacking model using Catboost and Bagging classifiers and Logistic Regression as meta model}
    \label{stack_model}
\end{figure}

In order to predict the exposure of organizations to hacking breaches, we build several models that leverage the designed socio-technical postures. We make use of tree-based and linear-based models to find the appropriate ones that would yield the best predictive performance. Tree-based models are mainly ensemble models based on boosting or bagging techniques. Boosting models by definition consist of a sequence of weak learners such as basic decision tree or decision stamps (i.e. decision trees with one level of depth) in a way that each learner is trained on the whole dataset to compensate the weaknesses of its predecessor. In the present study, we use three boosting models, namely, CatBoost \cite{prokhorenkova2018catboost},  LightGBM \cite{ke2017lightgbm} and XGBoost \cite{chen2016xgboost}. As for bagging, different samples are generated from the dataset with replacement, where each learner is trained on that sample, and the final result of a bagging model is obtained by averaging the outcomes of each learner. In our case, we use Random Forest and Bagging classifier \cite{scikit-learn}. In addition, we use logistic regression and support vector machine (SVM) \cite{scikit-learn} as linear-based models. Moreover, based on the best performing models, we build several stacked models combining models' predictions and using logistic regression as a meta-model to output the final predictions, as illustrated in figure \ref{stack_model}. Additionally, to explore the effectiveness of neural networks for our prediction task, we use TabNet \cite{Arik_Pfister_2021}, a neural network model based on attention mechanism. TabNet makes use of sequential attention which enables a feature selection process to focus on only important features at each decision step.

After preparing the data and selecting the predictive models, we split our dataset into a training set consisting of 70\% of the whole dataset, and use the remaining 30\% as a test set. In order to evaluate the performance of the selected models and provide a comprehensive evaluation, we use five metrics  including true positive rate (TPR), false positive rate (FPR), area under the ROC curve (AUC), f1-score (F1) and brier score defined as: $BS = \frac{1}{N}\sum_{i=1}^N (predictionProbability_i - actualClass_i)^2$, which evaluates how close the predictions probabilities are to the actual class. The lower the brier score is, the closer the predictions are to the actual classes.

\begin{table}[h]
    \centering
    \caption{Dataset split into 70\% training and 30\% testing after correcting labels with the fraction of breach and non-breach organizations}
    \label{train_test_split}
    \begin{tabular}{|l|c|c|}
    \hline
    Set & Breach (1) & No-Breach (0) \\ \hline
        Train set (70\%) & 29.35\%  & 70.65\% \\ \hline
        Test set (30\%)  & 31.55\%  & 68.44\% \\ \hline
    \end{tabular} 
    
\end{table}

\vspace{-2mm}

\section{Results and Discussion}

\subsection{Main Results}

\renewcommand\arraystretch{1.2}
\begin{table*}[h!]
\caption{STRisk prediction results on noise free data for technical features alone and all features combined. All features combined comprise technical measurements, Twitter indicators, sector and organization size }
\label{overall_results}
 \centering
 \begin{tabular}{|c|l|r|r|r|r|c|}
 \hline
 \multicolumn{2}{c|}{\cellcolor{gray!25}{}} & \multicolumn{5}{c|}{Metrics} \\
 \hline
Features  & \multicolumn{1}{c|}{Classifier} & \multicolumn{1}{c|}{TPR} & \multicolumn{1}{c|}{FPR} & \multicolumn{1}{c|}{AUC} & \multicolumn{1}{c|}{F1} & \multicolumn{1}{c|}{Brier score} \\
 \hline
\multirow{13}{*}{All features combined} &	CatBoost &	93.01\% &	3.67\% &	98.31\% &	0.89 &	0.046 \\
& 	LightGBM &	94.54\% &	5.31\% &	98.21\% &	0.90 &	0.043 \\
& 	XGBoost &	\textbf{95.45\% }&	6.04\% &	\textbf{98.43\% }&	\textbf{0.91} &	\textbf{0.042}  \\
& 	AdaBoost &	89.90\% &	12.91\% &	94.96\% &	0.80 &	0.243  \\
& 	Random Forest &	91.14\% &	4.90\% &	97.58\% &	0.85 &	0.065  \\
& 	Bagging classifier &	92.33\% &	4.71\% &	98.04\% &	0.87 &	0.062   \\
& 	Logistic Regression &	83.23\% &	30.63\% &	86.96\% &	0.63 &	0.13  \\
& 	SVM     &	82.05\% &	16.89\% &	87.37\% &	0.64 &	0.125  \\
& 	Naive Bayes  &	69.86\% &	16.67\% &	74.08\% &	0.02 &	0.195  \\
&   Satcking (Bagging + CatBoost) &	89.30\% &	1.17\% &	98.46\% &	0.83 &	0.127  \\
&   Satcking (Bagging + XGBoost)  &	91.95\% &	2.79\% &	98.45\% &	0.87 &	0.124  \\
&   Satcking (XGBoost + CatBoost) &	92.60\% &	\textbf{2.73\%} &	\textbf{98.46\%} &	0.88 &	0.114  \\
& TabNet & 75.19\% & 5.43\% & 91.69\% & 0.72 & 0.14 \\
 \hline
 \hline
 
\multirow{5}{*}{Technical features} &	CatBoost &	76.90\% & 	5.93\% & 	\textbf{86.33\%} & 	0.47 &	\textbf{0.12 }\\
& 	XGBoost &	\textbf{95.51\%} & 	44.97\% & 	85.92\% & 	0.69 &	0.13 \\
& 	RandomForest &	76.83\% & 	5.98\% & 	86.01\% & 	0.47 &	0.12 \\
& 	Logistic Regression &	71.09\% & 	24.32\% & 	83.90\% & 	0.14 &	0.17 \\
& 	Stacking (Bagging + CatBoost) &	74.93\% & 	\textbf{1.20\%} & 	85.97\% & 	0.38 &	0.18 \\ \hline
 \hline
{\makecell{Training on noisy labels}} & CatBoost & 94.79\% &	12.16\% &	97.71\% & 	0.83 & 	0.048 \\ 
 \hline
 \end{tabular}
 \end{table*}
\begin{table}
    \centering
    \caption{Comparison with prior work}
    \label{Comparison_tab}
    \begin{tabular}{|c|c|c|c|c|}
    \hline
        Work & TPR & FPR & FNR & AUC \\ \hline
        Cloudy \cite{liu2015cloudy} & 90\% & 10\% & 10\% & - \\
        Risky business \cite{sarabi2016risky} & 90\% & 11\% & 10\% & - \\ RiskTeller \cite{bilge2017riskteller} & - & - & - & 95\%\\
        \makecell{STRisk (CatBoost) \\ \(Technical\)} & 76.9\% & 5.93\% & 23.1\% & 86.33\% \\
        \textbf{STRisk (XGBoost)} & \textbf{95.45\%} & \textbf{6.04\%} & \textbf{4.55\%} & \textbf{ 98.43\%} \\ \hline
    \end{tabular}
\end{table}

Based on the prediction methodology discussed in section (\ref{problem_formulation}), we run different predictive models and we report the obtained results in table \ref{overall_results}. XGBoost and CatBoost are the best performing classifiers as they achieve the best AUC scores of 98.43\% and 98.31\% respectively, as well as the best TPR. Stacking XGBoost and CatBoost gives the best AUC score but lowers the TPR by about 3 points. This makes XGBoost the best classifier achieving the highest TPR and AUC, and lowest brier score. A lower brier score means that XGBoost is confident in its predictions and yields probabilities close to the actual class. When optimizing the FPR, stacking bagging classifer and CatBoost is the best option to get the lowest FPR. As for linear-based models, they do not yield satisfying performances in comparison to tree-based models making the latter better at capturing the differences between the two classes. Most importantly, social features combined with technical measurements, sector, and organization size, make a significant improvement in the prediction results with over 12\% in AUC compared to using only technical characteristics as predictors.

As detailed in Table \ref{Comparison_tab} considering all metrics, STRisk using XGBoost model outperforms prior work \cite{bilge2017riskteller} with over 3\% in AUC score. In \cite{bilge2017riskteller} random forest was used as predictive model on internally measured data including machines' logs. Measuring internally the bad symptoms is insightful, since it provides an internal view of the machines and unveil symptoms that can not be observed externally. However, this could be misleading and may not generalize to other organizations. We also outperform prior work \cite{liu2015cloudy} with a gain of 5.45\% in TPR and reducing FPR with about 4\%. In \cite{liu2015cloudy} random forest was used as classifier on externally assessed indicators such as misconfigured BGP and DNS data, expired https certificates, blacklists, etc. This is in line with our approach, except that in addition to technical security posture, we consider the social dimension which improves the performance of our system. Therefore, STRisk outperforms both internally and externally based risk assessment approaches.

\subsection{Prediction of Future Events}

In order to show how STRisk performs in predicting future data breaches (i.e. in  forecasting mode), we conduct additional experiments where we consider prior data incidents for a period of three months and six months. To do so, we collect hacking breaches occurred between March-2020 and March-2021 and build their socio-technical postures following the same methodology described above, except that we consider only social reputation and technical misconfigurations prior to the incidents. In this experiment we consider only victim organizations (perfectly labeled sample) and assess the performance of STRisk in predicting the breaches with only features measured before each incident. Using the same models trained on the dataset presented in table \ref{train_test_split}, we generate predictions for the considered organizations. For simplicity, we use CatBoost model and for the evaluation we consider as metric the accuracy that is obtained by the number of correctly classified incidents divided by the total incidents in the new sample. Table \ref{forecasting_res} shows the obtained accuracy using socio-technical posture data of the three months and six months leading to the reported incident date. We obtain an accuracy of 42.85\% using socio-technical data of three months prior the incident, and an accuracy of 71.42\% when using six months data. This is a significant improvement and indicates that the model benefits from additional data prior to the incident to better predict the breach risk.


\begin{table}[h]
    \centering
    
    \caption{STRisk-based prediction of new incidents using data aggregated over 3 months and 6 months prior to the incident }
    \label{forecasting_res}
    \begin{tabular}{|c|c|c|}
    \hline
        Model & 3 months prior to incident & 6 months prior to incident\\ \hline
        CatBoost & 42.85\% & 71.42\% \\
        \hline
    \end{tabular}
\end{table}

\subsection{Feature importance and interpretability}

\begin{figure}
    \centering
    \includegraphics[width=\linewidth]{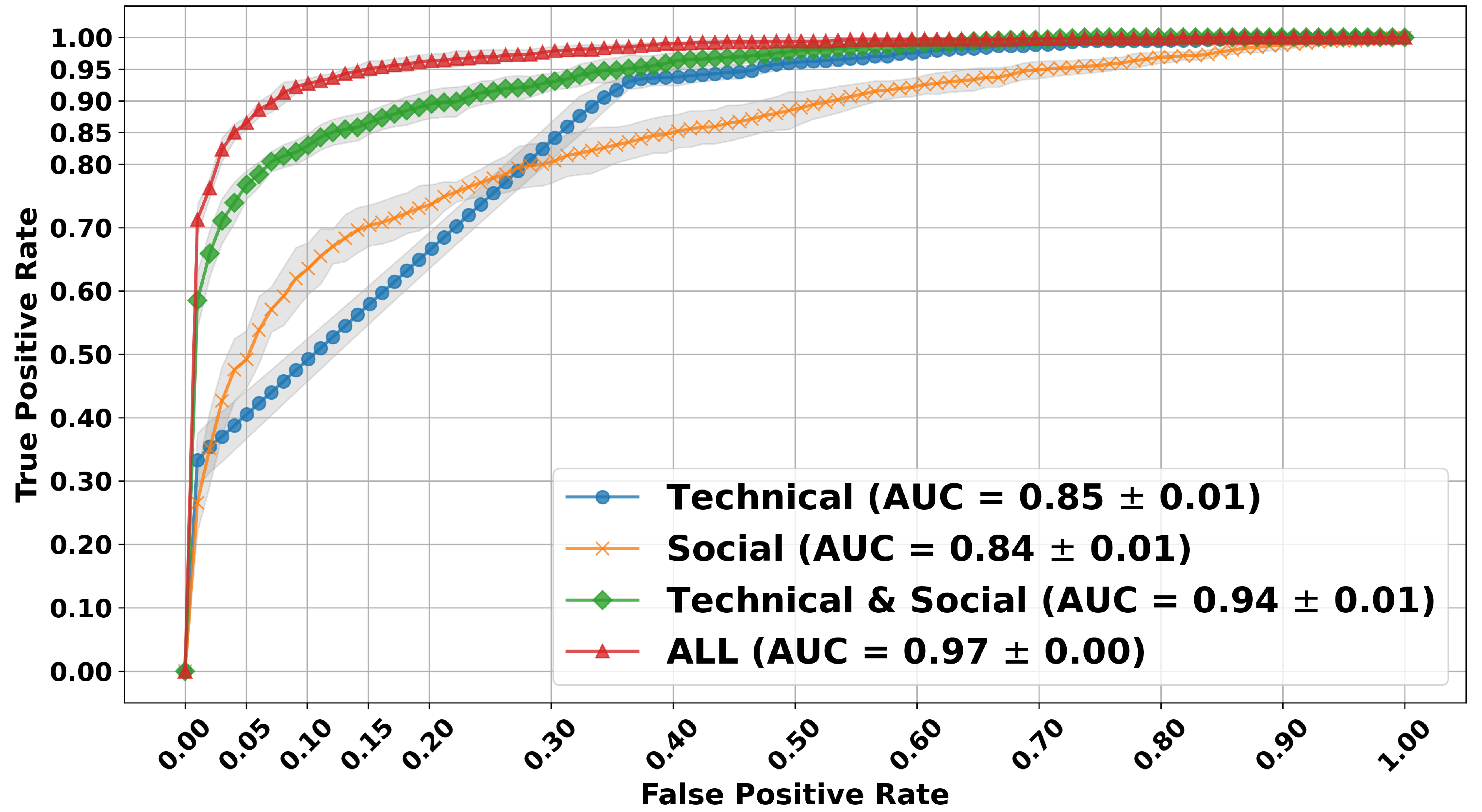}
    \caption{Separate predictions using each category of features alone Vs. overall predictions using all feature set that cover technical, social (Twitter), sector, and organization size}
    \label{separate_predictions}
\end{figure}

In order to study the importance of each feature in our feature set, we use the feature importance function returned by catboost which shows how changes in each feature affect the prediction value on average. As shown in table \ref{importances}, the feature importance returned by XGBoost emphasizes the significance of social features. These latter contribute to the model with over than 42\%, nearly the same share as the technical features contributing with 46\%. Besides, in order to assess the predictive power of each feature category, we build an XGBoost model for each category using 5-fold cross validation. By looking at the curves in figure \ref{separate_predictions} showing independent predictions by feature category, we observe that in one-to-one comparison, social features produce equal AUC score to using technical features and can be used alone for prediction which is an impressive achievement in data breaches prediction.

\begin{table}[h]
     \centering
     \caption{Feature importance by category}
     \label{importances}
     \begin{tabular}{|c|c|c|}
     \hline 
       Features category & Top features & Importance \\ \hline \hline
       \multirow{3}{*}{\makecell{Technical}} &  Open ports & \multirow{3}{*}{\makecell{46.8\%}} \\
       & Expired Certificates & \\
       & Darknet IPs & \\ \hline
       
       \multirow{3}{*}{\makecell{Twitter}} &  Likes ratio & \multirow{3}{*}{\makecell{42.3\%}} \\
       & retweet ratio & \\
       & Avg. polarity & \\ \hline
      
     Sector   & \cellcolor{gray!25}{} & 5.55\% \\ \hline
     Organization size & \cellcolor{gray!25}{} & 3.75\% \\ \hline
     \end{tabular}
 \end{table}

\begin{figure}[h]
     \centering
     \includegraphics[width=\linewidth]{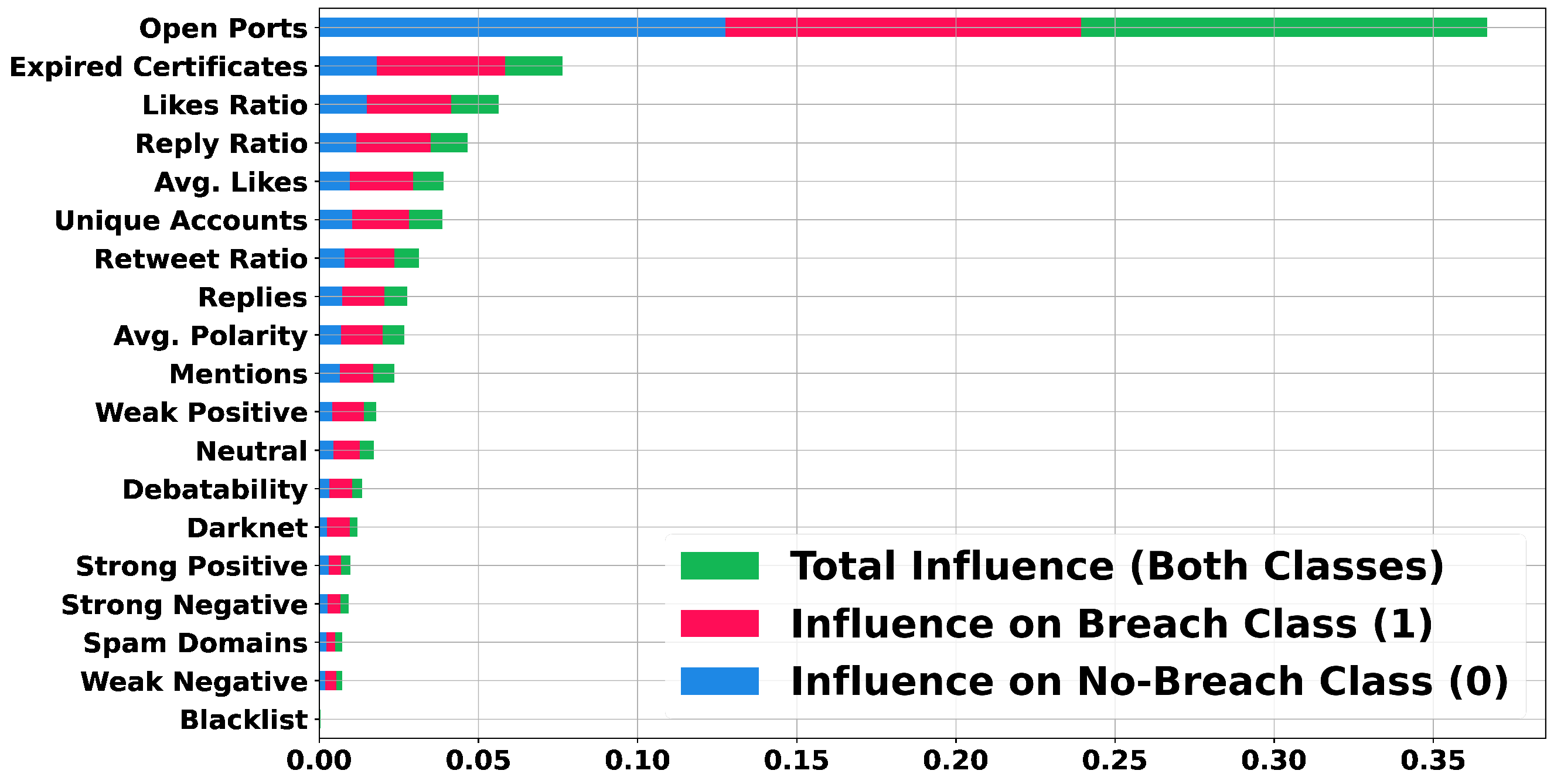}
     \caption{Per-class features interpretability on test set using SHAP values and CatBoost model}
     \label{shap_interpret} 
 \end{figure}

Additionally, to provide better and detailed interpretations on the most influential features on the model's predictions, we use SHapley Additive exPlanations (SHAP) method \cite{lundberg2017unified} which is a game theoretic approach to explain the predictions of any machine learning model. In our case since the best performing models are tree-based models, we use SHAP Tree Explainer \cite{lundberg2020local2global}. Fig. \ref{shap_interpret} shows the most influential features on the test set for both classes. The most influential features corroborates with the outcome of feature importance in Table \ref{importances} and indicates that open ports and expired certificates are the most important technical features. 

Diving into technical features, we find that the top-3 most contributing features are open ports, expired certificates, and IPs observed on the Darknet. These features could inform about various bad security symptoms. For example, if an organization have a critical open service/port such as remote services exposed to external exploitation, this may be leveraged by attackers to exfiltrate sensitive information or penetrate the system and persist in order to prepare for a damaging breach. A possible explanation for expired certificates is that they could increase phishing activities which may lead to identities theft. As for blacklisted IPs, this could be explained by an already infected machine and the infection could be the cause of the breach. Also, this could mean that the organization's hosts are engaged in malicious activities such as DDoS attacks. These bad indicators could not result directly in a breach, but would inform if an organization takes its security seriously. 

As for the most important social features, the top-3 most contributing social features are likes ratio, retweeting ratio and avg. polarity. Intuitively, these features would provide assessment of the organization's reputation on social media because of social influence and social propaganda. A possible explanation is that the values of these features could be high if the organization is known which could increase its attractiveness to attackers \cite{mezzour2017global}. These features could also reveal a bad posture if we suppose that a propaganda is launched on Twitter targeting a given organization. Likes and retweeting ratios for example, are significant in measuring whether the propaganda reaches a large portion of users. Therefore, a negative propaganda reaching a large portion of users would lead to a decrease in social reputation of the organization and finish on the radar of hackers and hacktivists. This insight agrees with the social influence theory that suggests that individuals' attitudes are influenced by others actors \cite{kelman1958compliance}, and joins the research on social cybersecurity that suggests that groups' opinions on social media can manipulate and shape the opinions of others \cite{carley2018social}. 

Additionally, adding Twitter features corrects the model's decision. By analyzing the profile of organizations that had experienced a breach and were not classified correctly when using only technical features, we find that adding Twitter features shifts the model's decision and corrects the prediction. The statistics of these organizations as shown in Table \ref{st_preds_stats}, indicate that almost all their technical features are null except a slight difference in open ports. This could mean that their security policy is good and it is hard for a predictive model based on technical profile to predict the existence of a breach. This shows the importance of adopting a socio-technical approach, which made the difference and reversed the model's decision to predict correctly victim organizations.  As for the few false positives that were incorrectly predicted as victims, Table \ref{false_positive_feats} shows a similar behaviour in terms of low technical misconfigurations as in Table \ref{st_preds_stats}, however the model could not predict the examples correctly using the social posture. The comparison between the two tables indicates share similar likes ratio, reply ratio, and avg. likes which are among the most influencing features based on Fig. \ref{shap_interpret}.

\vspace{-3mm}
 

\begin{table*}[]
    \centering
    \caption{Statistics of 86 organizations that had experienced a hacking breach and were correctly classified by STRisk when using socio-technical features but were not classified correctly when using only technical features. STD stands for standard deviation, Min for minimum value, Qu. for quartile, and Max for maximum value }
    \label{st_preds_stats}
    \resizebox{\textwidth}{!}{
    \begin{tabular}{lccccccccccccccccccc}
    \\ \hline
    & \makecell{Blacklist \\ratio} &  \makecell{Darknet \\ratio} &  \makecell{Expired \\certificates \\ratio} &  \makecell{Open \\ports \\ratio} &  \makecell{Spam \\domain \\ratio} &  \makecell{Unique \\ accounts} &  \makecell{Weak \\ neg.}  &  \makecell{Strong \\ neg.} &   \makecell{Weak \\ pos.} &  \makecell{Strong \\ pos.} &     Neutral &  Mentions & \# Replies &  \makecell{Retweet \\ ratio} & debatability & \makecell{Avg. \\ likes}  &  \makecell{Avg. \\ polarity}  &  \makecell{Likes \\ ratio}  & \makecell{Reply \\ ratio}  \\ \hline

    Mean  &         0.00 &             0.000004 &            3.547473e-07 &               1.804572 &             0.001900 &            274.87 &   36.33 &    5.91 &  177.25 &   45.11 &  247.72 &    472.01 &    141.06 &            1.52 &           4.47 &   1.85 &      0.13 &     2.18 &     0.04\\
    STD   &         0.00 &             0.000041 &            2.312510e-06 &               3.918620 &             0.017621  & 286.45 &   42.25 &   10.70 &  170.78 &   54.81 &  237.39 &    619.29 &    184.12 &            1.68 &           8.97 &   2.26 &      0.09 &     2.53&       0.43\\
    Min   &         0.00 &             0.00 &            0.00 &               0.00 &             0.00   &              0.00 &    0.00 &    0.00 &    0.00 &    0.00 &    0.00 &      0.00 &      0.00 &            0.00 &           0.00 &   0.00 &     0.00 &    0.00 &      0.00 \\
    Q1   &         0.00 &             0.00 &            0.00 &               0.00 &             0.00  &             10.00 &    0.00 &    0.00 &    8.00 &    0.00 &   11.00 &     12.50 &      2.00 &            0.60 &           0.32 &   0.14 &      0.07 &     0.47 &      0.01 \\
    Median   &         0.00 &             0.00 &            0.00 &               0.00 &             0.00  &            212.00 &   19.50 &    2.00 &  185.50 &   29.50 &  210.00&    223.00 &     63.00 &            1.15 &           3.00 &   1.33 &      0.14 &     1.87 &      0.12  \\
    Q3   &         0.00 &             0.00 &            0.00 &               0.987455 &             0.00 &            512.00 &   54.00 &    7.00 &  310.50 &   76.75 &  458.50 &    706.50 &    218.50 &            1.79 &           5.63 &   2.38 &      0.19 &     3.18 &      0.24  \\
    Max   &         0.00 &             0.000380 &            1.525879e-05 &              19.000000 &             0.163411   &   936 &  202 &   68 &  900 &  322 &  820 &   2785 &    901 &   10.33 &          80 &  13.07 &      0.49 &    13.07 &    0.80  \\ \hline
\end{tabular} }
\end{table*}

\begin{table*}[h]
    \centering
    \caption{Statistics of false positive, i.e. organizations incorrectly predicted by the model as victims}
    \label{false_positive_feats}
    \setlength{\tabcolsep}{2pt}
    \resizebox{\textwidth}{!}{
    \begin{tabular}{lrrrrrrrrrrrrrrrrrrrr}
    \hline
    {} & Org Size &  \makecell{Blacklist \\Ratio} &  \makecell{Darknet\\ Ratio} & \makecell{Expired \\Certificates\\ Ratio} &  \makecell{Open\\Ports\\ Ratio} &  \makecell{Spam \\Domains\\Ratio} &  \makecell{Unique \\ accounts} &  \makecell{Weak \\ neg.}  &  \makecell{Strong \\ neg.} &   \makecell{Weak \\ pos.} &  \makecell{Strong \\ pos.} &     Neutral &  Mentions & Replies &  \makecell{Retweet \\ ratio} & Debatability & \makecell{Avg. \\ likes}  &  \makecell{Avg. \\ polarity}  &  \makecell{Likes \\ ratio}  & \makecell{Reply \\ ratio } 
    \\ \hline
    Mean  &    129.30 &              0.0 &           0.00 &                        0.04 &              4.53 &                0.00 &           386.91 &          48.22 &             5.00 &         214.39 &            43.61 &   404.39 &    715.61 &         173.96 &            0.64 &          4.86 &       2.30 &          0.13 &         2.57 &           0.17 \\
    STD   &    254.27 &              0.0 &           0.00 &                        0.11 &             10.62 &                0.01 &           305.41 &          43.42 &             6.46 &         151.48 &            36.73 &   299.14 &    495.94 &         174.77 &            0.74 &          4.57 &       2.66 &          0.04 &         2.92 &           0.33 \\
    Min   &      8.00 &              0.0 &           0.00 &                        0.00 &              0.00 &                0.00 &             2.00 &           0.00 &             0.00 &           2.00 &             0.00 &     2.00 &      4.00 &           0.00 &           0.00 &          0.00 &       0.42 &          0.03 &        0.00 &          0.00 \\
    Q1   &     12.00 &              0.0 &           0.00 &                        0.00 &              0.00 &                0.00 &            58.50 &           5.50 &             0.00 &          39.00 &             8.00 &    83.50 &    161.50 &          18.00 &            0.26 &          1.54 &       1.01 &          0.10 &         0.89 &           0.06 \\
    Median   &     16.00 &              0.0 &           0.00 &                        0.00 &              0.00 &                0.00 &           381.00 &          47.00 &             1.00 &         277.00 &            41.00 &   501.00 &   1086.00 &          88.00 &            0.44 &          3.49 &       1.56 &          0.13 &         1.96 &           0.15 \\
    Q3   &    185.00 &              0.0 &           0.00 &                        0.00 &              3.33 &                0.00 &           703.50 &          89.00 &             8.00 &         323.00 &            67.50 &   625.00 &   1145.50 &         309.00 &            0.85 &          7.64 &       2.71 &          0.16 &         2.86 &           0.27 \\
    Max   &   1176.00 &              0.0 &           0.01 &                        0.50 &             45.38 &                0.03 &           846.00 &         116.00 &            20.00 &         444.00 &           116.00 &   914.00 &   1182.00 &         523.00 &            2.81 &         18.33 &      13.24 &          0.21 &        13.24 &           0.82 \\ \hline
    \end{tabular} } 
\end{table*}

\subsection{Discussion}

Orgnizations' network misconfigurations and malicious activities are certainly alarming indicators that may unveil a bad security policy. Therefore, assessing the security posture of an organization based on these indicators would provide an insightful view of risky organizations to experiencing hacking breach incidents. The results we obtained have further strengthened our hypothesis that network misconfigurations are good risk predictors achieving a prediction AUC score of 86.33\% which corroborate with what has been found in previous results \cite{liu2015cloudy}. In addition, our externally based approach yield a good performance against both externally-based approach \cite{liu2015cloudy} and internally-based approach \cite{bilge2017riskteller}. 

Arguably, building a risk assessment system relying only on technical indicators is not sufficient and may not reflect the real risk facing an organization. Thus, a better risk assessment system should take into account insights from other dimensions such as the social one. In fact, we show through STRisk that taking into account the social posture of an organization offers great predictive power and corrects bad predictions. This validates our hypothesis on the importance of these factors. This finding emphasizes the necessity of addressing breach incidents from socio-technical standpoint and should urge organizations to embrace a broader policy that glues social signals with technical indicators for better risk mitigation \cite{hills2015beyond}. 

Moreover, The predictions of STRisk using only data prior the data breach incident are promising and show that STRisk could benefit from a longer history of the socio-technical posture to provide accurate predictions. Thus, in terms of impact, STRisk could be beneficial for insurance companies in order to gain better visibility on the risk facing organizations, and to design better insurance plans for them. Also, STRisk predictions would help managers become well positioned to assess their risk to experience a hacking breach, and then to translate the technical security issues into mitigation strategies and insurance plans. Moreover, such predictions would help organizations suffering from late detection of breaches to reduce detection time by assessing their security and social postures. This would also contribute to help victim organizations to reduce the time spent containing a breach by identifying key misconfigurations and possible causes. 

\section{Limitations and Future Work}

Predicting organizations at risk of experiencing hacking breach is not trivial and many technical, social, political and other factors come into play. In this work, we considered both social and technical dimensions of hacking breaches and showed excellent results. for further insights, it would be interesting to consider additional technical data measured internally and externally. Other external data includes BGP, DNS and mail server, phishing, vulnerable applications data, to which we did not gain access. In addition, STRisk is trained on aggregated data over the considered period. Although it yields promising results in forecasting mode, it would interesting to train it in fully forecasting mode. Furthermore, although the negative sample is sampled randomly from the set of organizations affiliated to ARIN registry, it is difficult to distinguish between corporate networks and web servers \cite{sok_}. It is important to explore ways to filter out random web servers from the study to better quantify the contribution of the designed features to analyse data breach risk.

On the other hand, using Twitter as a rich source of information for gathering insights and sensing general opinion have showed a satisfactory predictive power. Besides, as features come from Twitter, there is evidence of the existence of bots and fake news that could alter the signals collected from Twitter \cite{shao2018spread, wang2018era}. Therefore, in the future, it would be interesting to focus on influential security experts and analysts, hacktivists and hackers which could provide more insights on the social reputation of an organization and the factors that could lead to a hacking breach. Also, in future work, it would be interesting to explore other social factors internally measured such as how much an organization spend on its security, as well as darkweb discussions as an additional social source of information on breaches and possible attack campaigns being prepared. 

Additionally, in this study we considered data breaches caused by hacking activities from external measurements. As future work, it would be interesting to consider all types of breaches and measure the bad symptoms by both external and internal measurements.

\section{Conclusion}

In this paper, we present STRisk, a socio-technical system for assessing hacking breach risk, addressing the problem of unreported breaches, and predicting organizations' exposure to hacking breaches. To evaluate the effectiveness of our approach, we study over 3800 organizations in the United States and design a socio-technical profile for each organization. We extend the technical risk assessment approach of measuring misconfigurations by bringing into play the social dimension consisting of signals from Twitter. By aggregating externally measured social and technical indicators, we achieve a prediction performance that exceeds 98\% AUC score, outperforming prior work. In addition, STRisk yields promising results when operating in forecasting mode using data prior the incident. The results show a significant contribution of social indicators that improve the AUC score by 12\% compared to using the technical posture alone. Furthermore, the fine-grained analysis of the features indicates that open ports and expired certificates are the most significant technical features. As for the social dimension, spreadability and agreeability are the most important Twitter features. Additionally, we find that the incorporation of Twitter features allows to predict victim organizations that are not correctly predicted using technical posture alone. With such findings, we emphasize the importance of considering a broader security policy taking into account social signals in addition to technical misconfigurations for a better risk assessment and mitigation. 

\section*{Acknowledgment}

We would like to thank our partners Jérôme François, Abdelkader Lahmadi, and Frederic Beck from the French Institute for Research in Computer Science and Automation (INRIA) for providing us with the darknet dataset.

\ifCLASSOPTIONcaptionsoff
  \newpage
\fi

\bibliographystyle{IEEEtran}
\bibliography{IEEEabrv, Bibliography}

\end{document}